\documentclass[9pt,journal]{IEEEtran}
\usepackage[utf8]{inputenc}
\usepackage{amsmath}
\usepackage{geometry}
\usepackage{booktabs} 
\usepackage{tabularx}  
\usepackage{graphicx}
\usepackage{float} 
\usepackage{array, makecell, ragged2e}

\usepackage{multirow}
\usepackage{longtable}
\usepackage{stfloats} 
\begin{document}
\title{Fetal Sleep: A Cross-Species Review of Physiology, Measurement, and Classification}
\author{%
Weitao~Tang\thanks{W. Tang is with the Department of Electrical and Computer Systems Engineering, Monash University, Melbourne, Australia.},
Johann~Vargas-Calixto\thanks{J. Vargas-Calixto is with the Department of Biomedical Informatics, Emory University, Atlanta, USA.},
Nasim~Katebi\thanks{N. Katebi is with the Department of Biomedical Informatics, Emory University, Atlanta, USA.},
Robert~Galinsky\thanks{R. Galinsky is with the Ritchie Centre, Hudson Institute of Medical Research, and the Department of Obstetrics and Gynaecology, Monash University, Melbourne, Australia.}, 
Gari~D.~Clifford\thanks{G. D. Clifford is with the Department of Biomedical Informatics, Emory University, Atlanta, USA, and also with the Department of Biomedical Engineering, Georgia Institute of Technology, Atlanta, USA.}, 
Faezeh~Marzbanrad\thanks{F.~Marzbanrad is with the Department of Electrical and Computer Systems Engineering, Monash University, Melbourne, Australia.}
\thanks{Research reported in this publication was supported in part by the National Institutes of Health through the Fogarty International Center and the Eunice Kennedy Shriver National Institute of Child Health and Human Development (Grant R01HD110480), by the Google.org AI for the Global Goals Impact Challenge Award, and by the National Health and Medical Research Council (NHMRC) of Australia (Grants 1124493 \& 1164954). N.K. is partially supported by a PREHS-SEED award (Grant K12ES033593).}%
}

\maketitle

\begin{abstract}
Fetal sleep is a relatively underexplored yet vital aspect of prenatal neurodevelopment. Understanding fetal sleep patterns could provide insights into early brain maturation and help clinicians detect signs of neurological compromise that arise due to fetal hypoxia or fetal growth restriction. This review synthesizes over eight decades of research on the physiological characteristics, ontogeny, and regulation of fetal sleep. We compare sleep-state patterns in humans and large animal models, highlighting species-specific differences and the presence of sleep-state analogs. We review both invasive techniques in animals and non-invasive modalities in humans. Computational methods for sleep-state classification are also examined, including rule-based approaches (with and without clustering-based preprocessing) and state-of-the-art deep learning techniques. Finally, we discuss how intrauterine conditions such as hypoxia and fetal growth restriction can disrupt fetal sleep. This review provides a comprehensive foundation for the development of objective, multimodal, and non-invasive fetal sleep monitoring technologies to support early diagnosis and intervention in prenatal care.
\end{abstract}

\begin{IEEEkeywords}
Fetal Sleep, Fetal Behavioral States, Fetal Monitoring, Sleep Classification, Neurodevelopment
\end{IEEEkeywords}
\section{Introduction}
Sleep plays a crucial role in brain development, synaptic plasticity, and metabolic regulation~\cite{mirmiran_development_2003,cerritelli_review_2021,luu_continuity_2023}. In neonates and infants, consolidated sleep–wake cycles have been shown to support neurodevelopmental milestones and long-term cognitive outcomes \cite{mirmiran_function_1995,mirmiran2003development,graven_sleep_2008,de_beritto_newborn_2020}. However, the nature and function of sleep before birth—during fetal life—remain poorly understood. This lack of understanding poses a significant gap in prenatal care, as fetal sleep may reflect underlying brain maturation and help identify early signs of neurological compromise or neurodevelopmental disorders. Developing objective and non-invasive fetal sleep monitoring tools could empower clinicians to assess neurodevelopmental trajectories, detect early signs of complications such as antepartum hypoxia or fetal growth restriction, and implement timely interventions to improve perinatal outcomes.

Although adult human sleep has been well studied over the past decades, the origins of sleep and circadian rhythm research date back to 1729. Jean-Jacques d’Ortous de Mairan observed that the leaves of Mimosa pudica continued their daily opening and closing cycles even in constant darkness~\cite{mairan1729observation}, providing the first scientific evidence of an endogenous biological clock. Building upon this, early physiological studies in humans were initiated by John Davy, who in 1845 systematically measured body temperature fluctuations across the sleep-wake cycle and suggested a relationship between temperature rhythms and rest~\cite{davy1845temperature}. 

Research on brain activity during sleep dates back nearly a century. In 1924, German psychiatrist Hans Berger successfully recorded the first human electroencephalogram (EEG), he published his landmark findings in 1929, providing the first objective evidence of brain wave activity during sleep and wakefulness~\cite{berger1929elektroenkephalogramm}. Loomis et al. recorded electrical activity from the human cerebral cortex and reported rhythmic variations in brain potentials influenced by mental activity, external stimuli, emotional states, and sleep in their 1936 study, “Electrical Potentials of the Human Brain”~\cite{loomis1936electrical}. However, systematic sleep studies can be considered to have begun in the 1950s, when Kleitman and Aserinsky discovered rapid eye movement (REM) sleep in humans \cite{aserinsky1953regularly}. Their 1953 landmark study demonstrated that REM episodes were associated with vivid dreams, marking the first clear differentiation between REM and  non-rapid eye movement (NREM) sleep. By 1957, further studies established cyclic variations in EEG during sleep, confirming that REM sleep recurs in predictable cycles throughout the night \cite{dement1957cyclic}. Stemming from the seminal work of Rechtschaffen and Kales~\cite{Rechtschaffen1968} and others in the 1960s, sleep staging criteria became standardized in adults. However, the earlier we look into human development, the less we understand about sleep. In particular, fetal sleep remains one of the least understood stages due to significant technical and ethical challenges associated with studying the developing brain in human fetuses. As a result, there have been disagreements regarding how to classify fetal neurobehavioral states. Nevertheless, as we discuss in this article, these states phenotypically display characteristics of REM and NREM sleep, and are therefore best referred to as fetal sleep states, which can provide valuable insights into neurodevelopment.

In parallel, fetal sleep studies emerged during the 1950s and 1960s. Early research primarily focused on fetal heart rate patterns and behavioral states, laying the groundwork for later investigations. In 1967, studies focusing on the relationship of intrauterine fetal activity to maternal sleep \cite{sterman_relationship_1967} and evidence of fetal-sleep cycles \cite{petre-quadens_sleep_1967} examined fetal movements in relation to maternal sleep, suggesting that distinct sleep cycles may exist in utero. The 1970s marked a significant shift, as technological advancements allowed for more precise monitoring of fetal brain activity in large animal models such as sheep and calves. A pivotal advance came in 1971, when Ruckebusch systematically recorded alternating high-voltage slow activity and low-voltage fast activity in fetal lambs and calves, associating high-voltage slow activity with NREM sleep and low-voltage fast activity with REM sleep or alert wakefulness~\cite{ruckebusch1971ecog}. His work laid the foundation for interpreting fetal EEG activity in relation to behavioral states. Building upon this, a 1974 study confirmed the presence of EEG-based fetal sleep stages and linked them to cardiovascular regulation in fetal sheep~\cite{mann_fetal_1974}. Further validation came in 1977, as alternating high-voltage slow activity and low-voltage fast activity states were again observed in fetal lambs, reinforcing their similarity to neonatal and adult sleep patterns~\cite{ruckebusch_sleep_1977}. Finally, a 1979 study on human fetuses identified short-term cyclic patterns of fetal activity \cite{granat_short-term_1979}, supporting the hypothesis that early sleep-wake cycles emerge before birth.

A major conceptual breakthrough occurred in 1982, when Nijhuis et al. introduced the fetal behavioral states (FBS) framework~\cite{nijhuis_are_1982}. This classification system defined distinct fetal states based on physiological and behavioral markers, drawing parallels between fetal and neonatal behavioral state patterns~\cite{nijhuis_are_1982}. Nijhuis further expanded the framework in a 1986 study~\cite{nijhuis_behavioural_1986}, providing a more detailed characterization of these states and their relevance to neurodevelopment and clinical assessments~\cite{nijhuis_behavioural_1986}. These findings confirmed that human fetuses exhibit four behavioral states (1F to 4F), analogous to neonatal states~\cite{nijhuis_are_1982, nijhuis_behavioural_1986}. States 1F and 2F correspond to quiet (NREM-like) and active (REM-like) sleep, respectively, while 3F is characterized by continuous eye movements without body movement, and 4F by vigorous body activity with unstable heart rate~\cite{nijhuis_behavioural_1986}. Fetal breathing, heart rate, and movement patterns were shown to serve as key indicators of neurodevelopment~\cite{nijhuis_behavioural_1986}. These insights reinforced the potential of FBS as a clinical tool for assessing fetal brain function and detecting potential developmental abnormalities~\cite{nijhuis_behavioural_1986}.

Despite extensive behavioral classifications such as the FBS framework~\cite{nijhuis_are_1982}, the presence of true wakefulness in the fetus remains controversial. In sheep, baboons, and humans alike, episodes of increased activity or arousal-like features often lack the sustained neural and behavioral markers associated with postnatal wakefulness~\cite{mellor_importance_2005, rigatto1986fetal, stark2013breathing}. Some have suggested that the apparent wakefulness state reflects a transitional phase between sleep states, similar to indeterminate sleep in the newborn~\cite{mirmiran_development_2003}, rather than true wakefulness~\cite{mellor_importance_2005}. This will be discussed in more detail later in the paper.


These foundational studies laid the groundwork for modern research, which continues to refine our understanding of fetal sleep across different species. However, despite eight decades of progress, there remains no comprehensive literature review that systematically integrates findings on fetal sleep physiology, measurement, classification, and disruption under pathological conditions. To address this gap, we synthesize and structure the state of the art in fetal sleep research. Specifically, we:

\begin{itemize}
    \item synthesize fetal sleep research across humans, sheep, and baboons, highlighting species-specific similarities and differences in sleep ontogeny and physiology;
    \item compare measurement modalities, ranging from invasive techniques in animal models to non-invasive approaches in human studies;
    \item review classification methods used to identify sleep states, spanning rule-based approaches—some of which incorporate multimodal signal integration—and recent advances in state-of-the-art deep learning; and
    \item examine how abnormal intrauterine conditions such as hypoxia and fetal growth restriction disrupt fetal sleep development and expression.
\end{itemize}

By bridging physiology, engineering, and clinical relevance, this review provides a foundation for future work in fetal neurodevelopment and the advancement of objective, fetal sleep-state monitoring tools.

\section{Background on Sleep}
\subsection{The Nature of Sleep}

Sleep is a fundamental biological process characterized by altered consciousness, reduced responsiveness to external stimuli, and decreased physical activity \cite{baranwal2023sleep}. It is broadly classified into two main phases: NREM and REM~\cite{patel2024physiology}. These phases alternate cyclically 
and are regulated by both circadian rhythms and homeostatic sleep pressure~\cite{patel2024physiology}. In adults, NREM sleep is subdivided into three stages: N1 (Stage 1) light sleep, N2 (Stage 2) intermediate sleep, and N3 (Stage 3) deep sleep or Slow-Wave Sleep (SWS). A typical adult sleep cycle progresses through N1 $\rightarrow$ N2 $\rightarrow$ N3 $\rightarrow$ N2 $\rightarrow$ REM~\cite{feinberg1979systematic}, repeating every 90 minutes for 4–6 cycles per night~\cite{memar2017novel}. REM sleep is characterized by rapid eye movements, skeletal muscle atonia, and dreaming. It is a paradoxical state due to its wake-like EEG (low amplitude, mixed frequency) combined with high arousal threshold and lack of muscle tone \cite{jones2004paradoxical,boissard2002rat,xi2001motor}. 

Ontogenetically, REM is the dominant state in neonates and gradually declines with age \cite{roffwarg1966ontogenetic, blumberg2005dynamics,blumberg2009form}. This high proportion suggests a crucial role in brain maturation. REM-related muscle twitches are now believed to serve as sensory feedback signals that help the developing brain refine motor maps and connectivity \cite{blumberg2020rem}. This ontogenetic role highlights the fundamental contribution of REM sleep to early-life brain development \cite{peever2017biology}.

As reviewed by Peever and Fuller (2017), REM sleep exhibits substantial interspecies diversity~\cite{peever2017biology}. While terrestrial mammals and birds generally display well-defined REM features, marine mammals such as dolphins appear to lack many classical characteristics, including rapid eye movements, cortical EEG desynchronization, skeletal muscle atonia~\cite{lyamin2008cetacean,mukhametov1987unihemispheric}. REM sleep may serve species-specific adaptive functions. Two main hypotheses have been proposed to explain interspecies variation in REM sleep. The energy allocation hypothesis posits that REM sleep conserves energy by suspending thermoregulation during this state, while the ontogenetic hypothesis suggests that REM sleep supports neural development and plasticity, particularly in altricial species~\cite{blumberg2020rem}.

\subsection{The Significance of Fetal Sleep}


Fetal sleep is essential for neurodevelopment, supporting the maturation of both the central (CNS) and autonomic nervous systems (ANS)~\cite{roffwarg1966ontogenetic,zizzo2020fetal,schneider2008human,graven_sleep_2008,hoyer2017monitoring,mulkey2018critical,samjeed2024fetal}. Cycling through distinct FBS promotes synaptogenesis, brain plasticity, and neuronal differentiation, contributing to postnatal cognitive and behavioral development~\cite{szeto_prenatal_1985,graven_sleep_2008}. Unlike postnatal sleep, fetal sleep is defined by physiological markers such as  heart rate variability (HRV), body movements, and EEG (in animal models), rather than behavioral reports~\cite{graven_sleep_2008,dipietro_fetal_2021,koome_ontogeny_2014}.

The emergence of distinct FBS such as quiet and active sleep during the third trimester has been interpreted as a sign of increasing functional brain complexity, driven by coordinated neural activity across developing circuits~\cite{scher2008ontogeny, graven_sleep_2008, van_den_bergh_fetal_2012}. This maturation is paralleled by the progressive development of the ANS, which plays a key role in supporting physiological regulation during fetal life~\cite{hoyer2017monitoring,mulkey2018critical,zizzo2020fetal}. In particular, the vagus nerve—a central component of the parasympathetic system—has been implicated in vital processes such as anti-inflammatory signaling and metabolic regulation across fetal, perinatal, and postnatal periods~\cite{herry2019vagal,bystrova2009novel}. From around 25 weeks of gestation, increasing vagal tone and myelination have been reported~\cite{mulkey2018critical, mulkey2019autonomic, schlatterer2022autonomic}, potentially contributing to the regulation of heart rate and state-dependent behaviors. As such, the emergence of distinguishable sleep states in the third trimester—detectable via patterns of fetal heart rate, eye movements, and body activity~\cite{nijhuis_are_1982}—may reflect both neural and autonomic maturation.

REM-like sleep is characterized by spontaneous fetal movements, including fetal breathing movements (FBM), which are thought to play a critical role in preparing vital systems for postnatal life~\cite{walker_fetal_2016}. When fetal movements are pharmacologically suppressed—such as through anesthesia or neuromuscular blockade—there is a marked reduction in oxygen consumption~\cite{rurak1983increased}, suggesting that fetal activity itself significantly contributes to metabolic demand. While this observation does not directly implicate REM sleep in energy conservation, it highlights the physiological cost of fetal activity and the potential adaptive role of sleep states in regulating energy expenditure.

Moreover, FBM—often observed during REM-like states—are believed to contribute to lung growth and maturation. Disruption of these movements, whether experimentally or in pathological conditions such as prolonged oligohydramnios, can result in pulmonary hypoplasia~\cite{fox1985fetal,gruenwald1957hypoplasia,wigglesworth1979effects,fewell1981effects,moessinger1983fetal,wigglesworth1976effects}. Although these disruptions are not exclusive to REM sleep, the strong association between FBM and REM-like states suggests that fetal sleep behavior may indirectly support pulmonary development. Altogether, these findings imply that FBS contribute not only to neurodevelopment, but also play a broader role in maintaining metabolic balance and promoting organ system maturation~\cite{van_den_bergh_fetal_2012}.

\subsection{Current Research Status and Gaps}

Advances in fetal monitoring techniques have enabled detailed characterization of sleep states using EEG, integrating electrocardiogram (ECG), and HRV analysis \cite{hoyer_fetal_2013, semeia_evaluation_2022}. Non-invasive methods such as fetal magnetoencephalography (FMEG) complement traditional measures by providing insights into fetal cortical activity \cite{stone_investigation_2017}. Current research has primarily focused on automated classification of fetal sleep states based on heart rate and movement patterns \cite{vairavan_computer-aided_2016, mercado_correlation_2024}. Studies have also explored EEG-based analysis of REM and NREM sleep across species \cite{koome_ontogeny_2014, myers_methods_nodate}. Additionally, researchers have investigated neural and autonomic markers of fetal brain maturation through spectral EEG and HRV analysis \cite{shaw_altered_2018, hoyer_fetal_2013}, and recent studies have also evaluated EEG spectral power and sleep state cycling to assess maturation after hypoxia–ischaemia in fetal sheep \cite{galinsky2023magnesium}. 

Despite these advancements, significant research gaps remain. The mechanisms linking fetal sleep disruptions to neurodevelopmental disorders are still poorly understood. Current measurement techniques are limited, relying on indirect and intermittent measurements fetal movement, heart rate variability and breathing to infer fetal sleep and behavioral states. Furthermore, cross-species comparisons lack standardization, making it difficult to generalize findings from animal models to human fetal development.

\section{Physiological Characteristics of Fetal Sleep}

\subsection{Historical Background: Discovery of Fetal Sleep States}

The scientific understanding of fetal sleep has evolved since the late 20th century. Nijhuis et al. first proposed a systematic classification of fetal behavioral states in the early 1980s based on physiological rhythms and movement patterns~\cite{nijhuis_are_1982}, followed by studies linking these states to fetal heart rate variability and motor activity~\cite{van_vliet_relationship_1985, nijhuis_behavioural_1986}. These findings suggested that sleep-wake regulation begins before birth. More recent advances in non-invasive fetal monitoring, including transabdominal fetal electrocardiography (FECG) and fetal magnetocardiography (FMCG), have confirmed the presence of sleep-like cycles in third-trimester human fetuses through heart rate variability analyses~\cite{pini_point_2020,mercado_correlation_2024}.



Cross-species investigations have further strengthened the concept of prenatal sleep organization. In fetal sheep, intrauterine polygraphic recordings—including EEG, electrooculography, and nuchal electromyography—demonstrated a developmental transition from disorganized to structured behavioral states around 115–120 days of gestation (approximately 80\% of term) \cite{szeto_prenatal_1985}. These recordings revealed alternating episodes of quiet sleep, REM sleep, and arousal, resembling adult-like sleep architecture and implying an early onset of sleep-wake regulation in utero. Similarly, studies in fetal baboons during the early 1990s identified EEG patterns indicative of both REM and NREM sleep, which closely resembled those seen in preterm human infants \cite{myers_quantitative_1993, stark_patterns_1994}. Together, these findings suggest that core aspects of fetal sleep-state differentiation are conserved across species with relatively mature central nervous systems at birth.

Together, human and animal studies show that fetal sleep states emerge prenatally and follow species-specific yet evolutionarily conserved patterns. The next section examines sleep-state classification across species.

\subsection{Fetal Sleep Cycle in Different Species}
Since direct recordings of fetal EEG and other neural activity are not feasible in humans, many fetal sleep studies rely on animal models \cite{walker_fetal_2016,schwab_investigation_2000}. Understanding the similarities and differences between species is therefore crucial for interpreting these findings and assessing their relevance to human development. 
Comparative studies across species provide insights into how fetal sleep develops and its evolutionary significance. 

Fetal sheep and baboons are widely used models for studying sleep state maturation due to their well-characterized sleep architecture and physiological similarities to humans \cite{schwab_time-variant_2006,schwab_investigation_2000,tournier_physiological_2022,lee_prostaglandin_2002,myers_methods_nodate,isler_local_2005,davidson_maternal_2011}. For example, fetal sheep exhibits a transition from disorganized to structured sleep patterns around 80\% gestation, mirroring human fetal sleep development at approximately 32-36 weeks gestation \cite{nijhuis_are_1982}. To better compare the sleep states of different species, we first take a deeper look at gestational length and birth weight, as shown in Table~\ref{tab:birth_weight_gestational_age}.

\begin{table}[t]
\caption{Comparison of Gestational Length and Birth Weight Across Species}
\label{tab:birth_weight_gestational_age}
\centering
\renewcommand{\arraystretch}{1}
\scriptsize
\begin{tabular}{|l|p{3cm}|p{2cm}|}
\hline
\textbf{Species (Fetus)} & \textbf{Gestation Length} & \textbf{Birth Weight (kg)} \\
\hline
Human  
& 37--41 weeks (259--294 days) \cite{pettker_antepartum_2018} 
& 3.02--3.80 \cite{stone_investigation_2017} \\
\hline
Sheep  
& 20--22 weeks (145--150 days) \cite{toubas_effect_1985,mann_fetal_1974,nijland_ovine_2000,szeto_prenatal_1985,lee_prostaglandin_2002,schwab_time-variant_2006} 
& 3.50--5 \cite{toubas_effect_1985,kelly2021interleukin} \\
\hline
Baboon 
& 25--26 weeks (175--180 days) \cite{stark_patterns_1994,myers_quantitative_1993} 
& 0.46--0.90 \cite{garland_fetal_nodate,myers_quantitative_1993} \\
\hline
\end{tabular}
\vspace{1mm}
\begin{flushleft}
\footnotesize{\textit{Note: Reported ranges are for illustrative reference and may vary across studies.}}
\end{flushleft}
\end{table}

From Table~\ref{tab:birth_weight_gestational_age}, we observe that sheep fetuses have a higher birth weight than human fetuses, while baboon fetuses are the lightest. Interestingly, this pattern does not follow the order of gestational length: human fetuses have the longest gestation (37–41 weeks), baboons fall in between (25–26 weeks), and sheep have the shortest (20–22 weeks). This dissociation between gestation length and birth weight suggests that longer gestation may not be solely for somatic growth. Instead, it may reflect species-specific neurodevelopmental priorities. In humans, for instance, the prolonged gestation supports a brain growth spurt that begins in mid-gestation and extends well into early postnatal life, enabling greater cortical and synaptic development \cite{dobbing1973quantitative}. This implies that cerebral complexity, rather than body weight, may better explain interspecies differences in gestation length—at least among medium-sized mammals. Moreover, human fetal birth weight is influenced by factors such as fetal sex \cite{dipietro2015studies}, maternal weight \cite{cerritelli_review_2021}, and gestational diabetes \cite{pettker_antepartum_2018}, while such influences have not been extensively studied in sheep or baboon models.

\begin{table*}[!b]
\caption{Comparison of Sleep State Differentiation Across Species}
\label{tab:sleep_species}
\centering
\renewcommand{\arraystretch}{1.2}
\scriptsize
\begin{tabular}{|l|p{3.5cm}|p{3.3cm}|p{4cm}|}
\hline
\textbf{Sleep State Differentiation} & \textbf{Human Fetus} & \textbf{Sheep Fetus} & \textbf{Baboon Fetus} \\
\hline
Initial Physiological Rhythmicity
& 80--90\% of gestation: FHR, eye, and body movements cycle independently; chance overlaps lack the synchrony and stability required for true behavioral states~\cite{nijhuis_are_1982}. 
& 79--83\% of gestation: 35.1 ± 2.5\% NREM, 52.7 ± 2.4\% REM, 11.2 ± 1.7\% Wake like activity were observed, marking the transition from disorganized to organized behavioral states.~\cite{szeto_prenatal_1985} 
& 78--86\% of gestation: EEG power and coherence cycles (1h) observed in fetal baboons, even in the absence of full behavioral state measures, may reflect early sleep-like rhythmicity \cite{isler_local_2005} \\
\hline
REM/NREM Differentiation 
& 90--95\% of gestation: In most fetuses, heart rate, movement, and eye activity are only partially synchronized, preventing reliable state classification.~\cite{nijhuis_are_1982} 
& 83--90\% of gestation: 38.4 ± 1.6\% NREM, 49.9 ± 1.7\% REM, 11.8 ± 1.6\% Wake like state~\cite{szeto_prenatal_1985} 
& 82--87\% of gestation: 20.9\% NREM, 58.3\% REM, 20.9\% Transition~\cite{grieve1994behavioral} \\
\hline
Emergence of Stable Behavioral States 
& 95\% of gestation: NREM: 32\% (range: 9–53.5\%)
REM: 42.5\% (range: 23–64\%)
Wake-like state (combined 3F and 4F): 14.75\% (range: 6.5–33\%)
No state identified: 11.5\% (range: 3–53.5\%)~\cite{nijhuis_are_1982} 
& 90--97\% of gestation: 38.2 ± 1.8\% NREM, 46.1 ± 2.0\% REM, 15.1 ± 1.9\% Wake like state~\cite{szeto_prenatal_1985} 
& 82--87\% of gestation: 36.8\% NREM, 63.2\% REM~\cite{stark1991characterization} \\
\hline
Full Maturation of Sleep Cycles 
& 99\% of gestation: NREM: 38\% (range: 24.5–52.5\%)
REM: 42.5\% (range: 22–73.5\%)
Wake-like state (combined 3F and 4F): 13.5\% (range: 2.5–38\%)
No state identified: 5\% (range: 0–26.5\%)~\cite{nijhuis_are_1982}
& 97--99\% of gestation: 43.8 ± 2.5\% NREM, 37.7 ± 2.8\% REM, 18.3 ± 2.9\% Wake like state \cite{szeto_prenatal_1985} 
& 73--90\% of gestation: 48\% NREM, 32\% REM, 20\% transition \cite{garland_fetal_nodate} \\
\hline
\end{tabular}

\vspace{1mm}
\begin{flushleft}
\end{flushleft}
\end{table*}

Comparison of sleep state development across species is summarized in Table~\ref{tab:sleep_species}. This table outlines key developmental milestones—ranging from the onset of physiological rhythmicity to the emergence and maturation of distinguishable sleep states—in humans, sheep, and baboons.

Among the three species, fetal sheep appear to exhibit organized sleep states earliest, with REM/NREM-like differentiation observable as early as 79\% of gestation. Fetal baboons also show distinct REM- and NREM-like cycling by 82–87\% of gestation, although data are limited to a narrow gestational window and lack a comprehensive developmental trajectory. In contrast, humans typically show consistent REM/NREM differentiation only after 90\% of gestation, suggesting a relatively delayed maturation process.

These cross-species comparisons provide insight into sleep state ontogeny; however, differences in methodology, available data, and sampling windows, particularly in the fetal baboon limit direct comparisons. The table should therefore be interpreted as an approximate alignment across species rather than a definitive staging framework.

\subsection{Comparison of Sleep-like States Across Species}

FBS are broadly classified into sleep-like states (REM and NREM), transitional or indeterminate states, and potential wakefulness. However, the nature and classification of fetal wakefulness remain a subject of debate. This section compares sleep patterns in fetal sheep, baboons, and humans based on available research.

\subsubsection{Sleep-like States}

All three species exhibit two primary sleep-like states: Quiet sleep and Active sleep. In fetal sheep, Quiet sleep is characterized by high-voltage low-frequency EEG, the absence of REMs, and variable nuchal muscle tone \cite{szeto_prenatal_1985,frank_complexity_2006}. Active sleep in fetal sheep is distinguished by a low-voltage high-frequency EEG pattern, the presence of REM, and a background of absent sustained nuchal muscle tone, often interspersed with phasic muscle contractions and breathing movements~\cite{szeto_prenatal_1985,clewlow1983changes}.

\begin{figure*}[htbp]
    \centering
    \includegraphics[width=\textwidth]{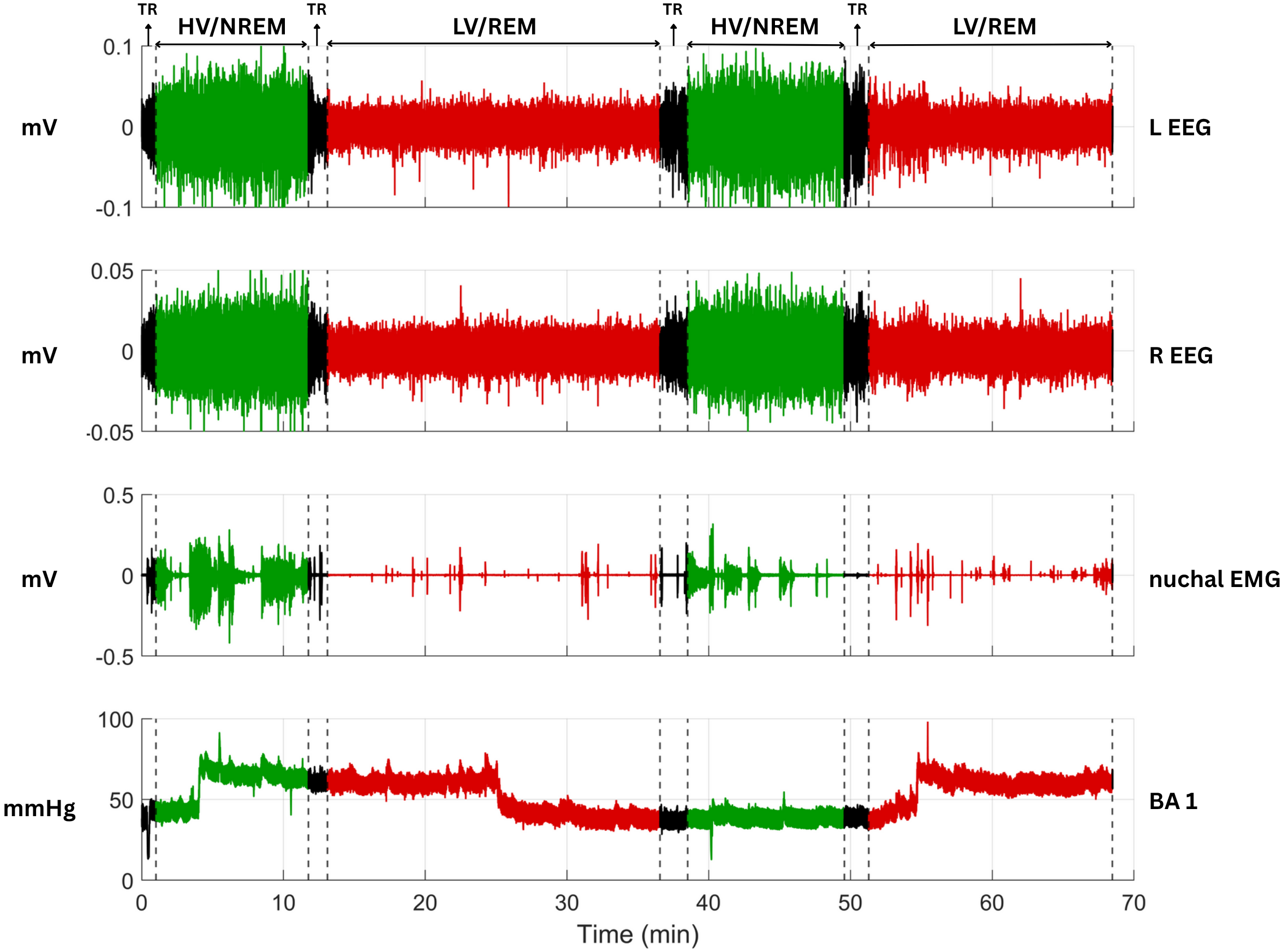}
    \caption{Representative physiological signals illustrating sleep states in fetal sheep~\cite{tang2024advancing}. NREM sleep is marked by high-voltage (HV), low-frequency EEG patterns recorded from both hemispheres (L EEG and R EEG), whereas REM sleep is characterized by low-voltage (LV), high-frequency EEG activity. TR represents an intermediate state between REM and NREM, capturing the dynamic shift from one state to the other. This state typically exhibits mixed EEG features that do not fully conform to either REM or NREM characteristics. Additional signals include nuchal EMG, obtained from electrodes implanted in the fetal neck muscles, which reflects muscle tone and fetal movements. BA 1 denotes the raw intra-balloon pressure signal, capturing both the balloon inflation pressure and the ambient amniotic pressure. All signals were collected from fetal sheep using chronic invasive instrumentation, including surgically implanted EEG and EMG electrodes and an intra-amniotic balloon catheter, enabling continuous in utero monitoring of physiological and neural activity.}
    \label{fig:label_presentation}
\end{figure*}

To better elucidate the physiological distinctions between NREM and REM sleep, we provide representative traces from our own fetal sheep recordings. These examples highlight characteristic differences in EEG, EMG, and intra-balloon pressure across behavioral states. As illustrated in Figure~\ref{fig:label_presentation}, these features are clearly distinguishable: green waveform segments correspond to quiet sleep (HV/NREM), red segments indicate active sleep (LV/REM), and black segments denote brief transition periods (TR).

Similarly, fetal baboons display quiet and active sleep states. Quiet sleep is associated with trace alternant, a pattern of intermittent bursts of high-voltage EEG activity also reported in other species such as fetal sheep, whereas active sleep exhibits an increased presence of high-frequency EEG components \cite{myers_quantitative_1993}. Fetal breathing is present in both sleep states but occurs more frequently in active sleep~\cite{stark_patterns_1994}, a pattern consistent across species. The organization of these states resembles that of fetal sheep and humans.

In humans, these two states are referred to as Quiet sleep (1F) and Active sleep (2F). Quiet sleep is characterized by infrequent fetal movement, stable fetal heart rate with low variability, and an absence of eye movements (EOG), akin to NREM sleep; whereas Active sleep, corresponding to REM sleep, involves frequent fetal body movements, continuous EOG, and a variable heart rate \cite{nijhuis_behavioural_1986,lowe_parturition_1996}.

\subsubsection{Sleep Transitions and Indeterminate Sleep}

All three species—sheep, baboons, and humans—exhibit transitional or indeterminate fetal sleep states, reflecting the developmental complexity of sleep organization.

In fetal sheep, these ambiguous states—often termed intermediate sleep—are neither clearly REM nor NREM. Mellor (2005) linked them to immature brain regulation \cite{mellor_importance_2005}. Quantitative EEG studies identified two spectral intermediates, falling between high-voltage slow activity and low-voltage fast activity, as exemplified by the TR highlighted in black in Figure~\ref{fig:label_presentation}, accounting for ~23\% of recording time \cite{mcnerney_automated_nodate}. Rao et al. (2009) defined indeterminate sleep as transitional or mismatched EEG/EOG periods, excluded if under 3 minutes \cite{rao2009behavioural}. Such states may reflect neural transitions key to sleep maturation \cite{ioffe_sleep_1980}.

Fetal baboons also show EEG evidence of graded transitions between quiet and active sleep. EEG-ratio analyses reveal sleep as a continuum, with up to 60\% of time spent in indeterminate states \cite{myers_quantitative_1993,garland_fetal_nodate}.

In humans, fetal sleep cycles last 70–90 minutes, with state transitions typically within 3 minutes \cite{visser1992studies,mulder1998acute}. Indeterminate states, defined by mismatched behavioral markers, occur in approximately 5–10\% of recordings near term \cite{nijhuis_are_1982}, suggesting lower prevalence compared to baboons.

\subsubsection{Fetal Arousal and Wakefulness}

The definition of arousal and wakefulness in the fetus remains contentious. In fetal sheep, brief periods of activity characterized by low-voltage electrocorticography (ECoG), EOG, and increased EMG activity have been interpreted as an aroused state \cite{szeto1992behavioral,crossley_suppression_1997,nicol1998effect,nicol2001effect}. Nevertheless, accumulating evidence suggests that such episodes may merely reflect transitional phases between sleep states, rather than true wakefulness \cite{mellor_importance_2005}. Direct observations of unanaesthetized fetal sheep provide further evidence against the existence of true wakefulness in utero. Rigatto et al. \cite{rigatto1986fetal} observed a fetal sheep through a Plexiglas window for 5,000 hours and found no signs of wakefulness, such as eye opening or coordinated head movements. This suggests that fetal sheep remain in sleep-like states throughout gestation without experiencing a state that would be comparable to postnatal wakefulness.

In fetal baboons, studies indicate that hiccups and gross fetal movements do not necessarily induce sleep-state transitions, implying that fetal activity does not always equate to wakefulness \cite{stark2013breathing}. Some researchers have suggested that fetal wakefulness, if present, is rare and might be a misclassification of state transitions \cite{myers_quantitative_1993}.

In human fetuses, state 3F (quiet awake) is rarely observed, and state 4F (active awake), though defined, is difficult to identify reliably due to obscured eye movements \cite{nijhuis_are_1982}. Some researchers suggest that fetal wakefulness may not exist in the same form as in neonates; episodes commonly interpreted as wakefulness could instead reflect transitions between sleep states \cite{mellor_importance_2005}. As inferred from electrophysiological evidence, EEG activity during such periods may resemble spontaneous sleep-state transitions rather than sustained wakefulness \cite{mellor_importance_2005}.

The ability to intentionally wake the fetus is not well established. In fetal sheep, external stimuli such as maternal hormone fluctuations influence fetal sleep patterns, but there is little evidence that these lead to wakefulness \cite{lee_prostaglandin_2002,mellor_importance_2005}. In fetal baboons, experimental challenges such as hypoxia and auditory stimuli have been proposed to investigate fetal state changes, but the extent to which these lead to true wakefulness is unclear \cite{stark2013breathing}. For human fetuses, vibroacoustic stimulation (VAS) has been studied as a method to elicit state transitions, but behavioral state organization remains largely resistant to external influences \cite{visser_effect_1993}. Nevertheless, early acoustic studies by Walker (1971) identified consistent intrauterine sound patterns originating from maternal cardiovascular activity, suggesting the fetal environment is shaped by rhythmic auditory input \cite{walker1971intrauterine}. Moreover, term fetuses have been shown to display differential heart rate responses to their mother’s voice compared to a stranger’s, indicating a capacity for auditory learning and in utero voice recognition \cite{kisilevsky2003effects}. Although factors such as maternal emotions, Braxton Hicks contractions, and uterine contractions during labor do not significantly alter fetal behavioral state patterns \cite{visser_effect_1993}, these findings imply that specific types of auditory stimulation may modulate fetal physiology without necessarily inducing full wakefulness. All three species spend the majority of their time in sleep-like states, suggesting that fetal wakefulness, if it exists, is actively suppressed. In fetal sheep, mechanisms such as prostaglandin-mediated regulation and hypoxia-induced depression of breathing contribute to the maintenance of prolonged sleep-like states \cite{lee_prostaglandin_2002,mellor_importance_2005}. Baboons exhibit a similar predominance of sleep-like states, with only brief transitions into undefined states \cite{stark_patterns_1994}. In humans, the fetal nervous system appears to be adapted for continuous sleep-like states, with developing neuronal circuits reinforcing these patterns \cite{reissland_fetal_2016,van_den_bergh_fetal_2012}.

Fetal sheep, baboons, and humans exhibit similar sleep-like states, though true wakefulness remains unclear. Transitional states are frequent, and sleep organization matures with gestation. Across species, sleep state dominates with little evidence of sustained wakefulness before birth.

\subsection{Maternal and External Factors}

Maternal physiology and environmental conditions have been shown to influence fetal sleep states \cite{stone_investigation_2017,wilson_maternal_2022}. Fetal sleep rhythms begin to develop in utero and are thought to be entrained by maternal melatonin and circadian cues \cite{rajendram_childhood_2017}. Various factors, including maternal sleep position, circadian rhythms, sleep disorders, and external stressors, contribute to the regulation of fetal brain activity and behavioral states. External stressors can also influence fetal sleep states. For example, hypoxia suppresses FBM and modifies EEG activity, a process likely mediated by elevated adenosine levels, which act as an inhibitory neuromodulator, and by increased neurosteroids such as allopregnanolone, which suppress neuronal excitation and protect the fetal brain \cite{efsa_panelon_animal_health_and_welfare_ahaw_animal_2017, yawno_neuroactive_2011}.

The position a mother adopts during sleep may affect FBS. Supine sleep is known to be associated with reduced uteroplacental perfusion, leading to fetal quiescence \cite{stone_investigation_2017}. This reduction may be due to altered maternal cardiac output and uteroplacental perfusion, which transiently affect oxygen and nutrient delivery to the fetus, potentially influencing fetal sleep patterns and activity levels \cite{kinsella1994supine, pirhonen1990uterine, jeffreys2006uterine}.
 Late stillbirth is independently related to the position women adopt during sleep \cite{stone_investigation_2017}. Vulnerable fetuses, who may already experience chronic hypoxia, have a reduced ability to adapt to maternal sleep position stressors \cite{stone_investigation_2017}. Another study found that passive maternal movements, such as rocking or swaying, can alter fetal heart rate and potentially the behavioral states, likely through activation of the vestibular system \cite{lecanuet2002fetal}.

Fetal sleep patterns are also closely linked to maternal circadian rhythms. Studies have shown that fetal sleep states align with maternal melatonin secretion and activity-rest cycles, indicating that maternal circadian rhythms play a role in regulating fetal brain activity \cite{mirmiran1996perinatal}. Furthermore, maternal sleep-disordered breathing (SDB), such as sleep apnea, becomes more common in the third trimester and can disrupt the intrauterine environment by inducing nocturnal hypoxia and heightened maternal autonomic activity. These changes have been associated with alterations in fetal physiological behaviors, including heart rate decelerations and reduced fetal breathing movements—both of which are key indicators of fetal sleep states. While the precise impact on long-term neurodevelopment remains uncertain, these findings suggest that maternal SDB may acutely influence fetal sleep regulation~\cite{pien2004sleep,sahin2008obstructive}.

In summary, maternal physiological and environmental factors have significant effects on fetal sleep states through mechanisms involving hemodynamics, hormonal regulation, and neural modulation. These factors collectively contribute to shaping fetal brain activity and behavioral states.

\begin{table*}[!t]
    \centering
    \caption{Comparison of Fetal Sleep Measurement Techniques}
    \label{tab:fetal_measurement_comparison}
    \renewcommand{\arraystretch}{1.4}
    \scriptsize
    \begin{tabular}{|p{2cm}|p{3.5cm}|p{4.2cm}|p{4.2cm}|}
        \hline
        \textbf{Measurement} & \textbf{Fetal Human (Non-invasive)} & \textbf{Fetal Sheep (Invasive)} & \textbf{Fetal Baboon (Invasive)} \\
        \hline
        EEG/ECoG & 
        Infeasible & 
        Implanted electrodes~\cite{koos_adenosine_2001,mcnerney_automated_nodate,hinman_cholinergic_1988} & 
        Dural electrodes~\cite{myers_methods_nodate,stark_patterns_1994,myers_quantitative_1993} \\
        \hline
        EOG & 
        \makecell[l]{Ultrasound imaging~\cite{han_fetal_2018,visser_effect_1993}} & 
        \makecell[l]{Canthus electrodes~\cite{hinman_cholinergic_1988,szeto_prenatal_1985,ioffe_sleep_1980}} & 
        \makecell[l]{Canthus electrodes~\cite{grieve1994behavioral,garland_fetal_nodate}} \\
        \hline
        EMG & 
        Infeasible & 
        \makecell[l]{Nuchal EMG~\cite{burchfield_cocaine_1990,nicol_effects_1999} \\
        Diaphragmatic EMG~\cite{ioffe_fetal_1993}} & 
        Not Commonly used \\
        \hline
        FHR & 
        \makecell[l]{CTG~\cite{pini_point_2020,vairavan_computer-aided_2016,alfirevic2017continuous} \\
        Abdominal ECG~\cite{karin_estimate_1993} \\
        Scalp ECG (Invasive)~\cite{van_laar_power_2009} \\
        FMCG~\cite{chiera2020heart,van2014fetal,lowery2008assessing,brandle_heart_2015}} & 
        \makecell[l]{Arterial pressure~\cite{burchfield_cocaine_1990,hasan_arterial_1992} \\
        ECG~\cite{jensen_role_2009}} & 
        ECG electrodes~\cite{garland_fetal_nodate} \\
        \hline
        FBM & 
        \makecell[l]{Ultrasound imaging~\cite{han_fetal_2018,visser_effect_1993}} & 
        \makecell[l]{Tracheal catheter~\cite{richardst_respiratory_nodate} \\
        Diaphragm EMG~\cite{harding_ingestion_1984} \\
        Laryngeal EMG (PCA)~\cite{harding_ingestion_1984}} & 
        \makecell[l]{Tracheal catheter + amniotic subtraction\\~\cite{stark_patterns_1994,stark2013breathing,garland_fetal_nodate}} \\
        \hline
        Body Movements & 
        \makecell[l]{Actocardiogram~\cite{dipietro2015studies,brandle_heart_2015}} & 
        \makecell[l]{Limb EMG~\cite{szeto_prenatal_1985,natale_measurement_1981} \\
        Ultrasound~\cite{poore1980chest}} & 
        Not Commonly used \\
        \hline
    \end{tabular}
\end{table*}


\section{Acquisition of Physiological Signals in the Fetus}

To understand fetal behavioral and sleep states, researchers have relied on both non-invasive technologies suitable for human fetuses and invasive modalities enabled by animal models such as fetal sheep and baboons. Table~\ref{tab:fetal_measurement_comparison} summarizes the key measurement techniques across species. In the remainder of this section, we describe each modality in more detail, with a focus on signal types, acquisition methods, and the physiological information derived.

\subsection{Non-Invasive Technologies for Human Fetus}

In human fetal research, ethical and technical limitations necessitate the use of non-invasive techniques. These technologies prioritize safety, cost-effectiveness, and practicality, while attempting to capture physiological signals linked to fetal behavioral states.

\begin{itemize}
\item \textbf{Cardiotocography (CTG)}: A widely used method employing 1D Doppler ultrasound to monitor fetal heart rate (FHR) and uterine contractions through the maternal abdomen~\cite{pini_point_2020,vairavan_computer-aided_2016,alfirevic2017continuous}. CTG is low-cost and non-invasive~\cite{guo_prenatal_2021}, but it provides only a smoothed heart rate estimate rather than beat-to-beat intervals, limiting detailed HRV analysis~\cite{hirsch1995heart,lange_heart_2009}.

\item \textbf{FECG}: Electrodes on the maternal abdomen record fetal cardiac signals, though maternal ECG interference often degrades signal quality~\cite{karin_estimate_1993}. A scalp electrode applied intrapartum offers improved fidelity but is invasive and limited to labor~\cite{van_laar_power_2009,lange_heart_2009}.

\item \textbf{FMCG}: A high-resolution modality that uses superconducting quantum interference devices sensors to detect fetal cardiac magnetic fields through the maternal abdomen~\cite{chiera2020heart,van2014fetal,lowery2008assessing,brandle_heart_2015}. It offers millisecond temporal resolution and is reported to be less susceptible to artifacts than FECG~\cite{peters2001monitoring}, but it is expensive and technically demanding.

\item \textbf{Ultrasound Imaging}: Ultrasound is used to monitor fetal body and eye movements, amniotic fluid volume, breathing, and muscle tone~\cite{han_fetal_2018,visser_effect_1993}.

\item \textbf{Actocardiography}: Actocardiography combines Doppler-derived FHR and movement data~\cite{dipietro2015studies,brandle_heart_2015}, enabling richer behavioral state characterization~\cite{mercado_correlation_2024}.

\end{itemize}

\subsection{Invasive Technologies in Animal Models}

\begin{table*}[t]
\centering
\caption{Comparison of EEG and FHR Frequency Band Definitions Across Species}
\label{tab:freq_bands}
\renewcommand{\arraystretch}{1.3}
\scriptsize
\begin{tabular}{|l|p{3.5cm}|p{3.5cm}|p{3.5cm}|}
\hline
\textbf{Signal Type} & \textbf{Fetal Human} & \textbf{Fetal Sheep} & \textbf{Fetal Baboon} \\
\hline
EEG 
& Not available in fetal human studies
& Delta (0--3.9 Hz), Theta (4--7.9 Hz), Alpha (8--12.9 Hz), Beta (13--22 Hz)~\cite{lear_dysmaturation_2024,tang2024advancing}
& Delta (1--4 Hz), Theta (4--7 Hz), Alpha (8--12 Hz), Beta1 (14--18 Hz), Beta2 (22--29 Hz)~\cite{isler_local_2005} \\
\hline
FHR 
& VLF (0.02--0.08 Hz), LF (0.08--0.2 Hz), Intermediate (0.2--0.4 Hz), HF (0.4--1.7 Hz)~\cite{david2007estimate,gustafson2011characterization,gustafson2012fetal}
& VLF (0--0.04 Hz), LF (0.04--0.15 Hz), HF (0.15--0.4 Hz)~\cite{min2002power,van2009power,koome2014quantifying}
& LF (0.05--0.2 Hz), HF (0.5--2.0 Hz)~\cite{garland_fetal_nodate} \\
\hline
\end{tabular}
\end{table*}

Several fetal monitoring techniques used in humans, including FECG and FMCG, are also employed in animal models. In particular, fetal sheep and baboons enable the use of invasive methods that offer high-resolution, direct physiological measurements. These modalities facilitate a more granular analysis of fetal sleep and behavior, including electrocortical activity, eye movements, respiration, muscle tone, and cardiovascular dynamics.

\begin{itemize}
\item \textbf{EEG/ECoG}: Electrodes implanted on or beneath the fetal skull record electrocortical activity. In sheep, stainless-steel screws and solder-ball electrodes are used~\cite{koos_adenosine_2001,mcnerney_automated_nodate,hinman_cholinergic_1988,kelly2023progressive}; in baboons, electrodes are placed on the dura mater~\cite{myers_methods_nodate,stark_patterns_1994}. 

 Signals are filtered (0.1–40 Hz or up to 100 Hz) and digitized at 50–200 Hz~\cite{hasan_arterial_1992,isler_local_2005}. Sleep states are differentiated by power spectral analysis: quiet sleep shows 1–4 Hz bursts (Trace Alternans), while active sleep exhibits elevated 12–24 Hz power~\cite{myers_quantitative_1993}. Common measures include spectral edge frequency (SEF) and EEG-ratio (0.03–0.2 Hz vs. 12–24 Hz)~\cite{isler_local_2005,nijland_ovine_2000,ioffe_ecog_1984}.

\item \textbf{EMG}: captures muscle activity via implanted electrodes and is used in fetal sheep to assess neuromuscular and respiratory activity. Limb EMG detects gross body movements through electrodes in muscles such as the quadriceps and triceps~\cite{szeto_prenatal_1985,natale_measurement_1981}, while nuchal EMG assesses muscle tone and sleep state transitions via electrodes in neck muscles~\cite{burchfield_cocaine_1990,nicol_effects_1999}. Diaphragmatic EMG reflects respiratory-related muscle activity and is used to detect FBM~\cite{ioffe_fetal_1993}.

\item \textbf{EOG}: detects eye movements through electrodes implanted near the orbits (sheep)~\cite{hinman_cholinergic_1988} or subcutaneously around the eye (baboons)~\cite{grieve1994behavioral}. 

\item \textbf{Respiratory Activity Monitoring}: FBM are monitored invasively using pressure catheters or EMG electrodes. In sheep, a pressure catheter is placed in the fetal trachea to detect intrathoracic fluid shifts~\cite{richardst_respiratory_nodate}, while EMG electrodes can be sewn into respiratory muscles such as the diaphragm or posterior cricoarytenoid~\cite{harding_ingestion_1984}. In baboons, tracheal and amniotic fluid pressures are measured with separate catheters to isolate breathing activity—a method also commonly used in fetal sheep studies~\cite{stark_patterns_1994,stark2013breathing,garland_fetal_nodate}. 

Breathing is typically intermittent and linked to REM-like sleep, characterized by low-voltage ECoG~\cite{richardst_respiratory_nodate}. These fetal breathing movements are essential not only for lung development but also for training the neural circuits that control respiration and for strengthening respiratory muscles in preparation for breathing after birth. Data are digitized at rates such as 25 Hz for waveform analysis~\cite{stark_patterns_1994}.

\item \textbf{Cardiovascular Signal Acquisition}: Fetal heart rate is derived from ECG, arterial pressure signals, or Doppler flow probes secured onto major arteries. In sheep, ECG electrodes are implanted on the chest~\cite{jensen_role_2009} or pressure waveforms are recorded via catheters in fetal arteries ~\cite{burchfield_cocaine_1990,hasan_arterial_1992}. In baboons, ECG leads with silver solder balls are fixed beneath the skin over the precordium~\cite{garland_fetal_nodate,myers_methods_nodate}. 

\end{itemize}


\subsection{Comparison of EEG and FHR Frequency Bands}
Table~\ref{tab:freq_bands} compares EEG and FHR frequency band definitions across fetal human, sheep, and baboon studies. This table highlights a key challenge in cross-species comparisons: the frequency band boundaries, especially for EEG rhythms, vary considerably due to both biological differences and species-specific research conventions. For instance, delta and theta bands in fetal sheep span wider frequency ranges than in baboons. Similarly, the definition of FHR bands such as very low frequency (VLF) and high frequency (HF) also differs between species, which complicates the translation of findings from animal models to human contexts. Recognizing these inconsistencies is essential for interpreting spectral analyses and designing cross-species comparative studies.

\section{Automatic Classification of Fetal Sleep}
Fetal sleep classification has been explored using rule-based and deep learning approaches. Rule-based methods rely on expert-defined thresholds and logic rules, sometimes supported by clustering-based preprocessing such as K-means. In contrast, deep learning enables end-to-end, data-driven modeling from raw physiological signals. Table~\ref{tab:combined_fbs_comparison} summarizes key distinctions across these approaches. The following subsections detail each category, including recent developments in multimodal signal integration.

\begin{table*}[t]
\centering
\caption{Comparison of FBS Classification Studies}
\label{tab:combined_fbs_comparison}
\renewcommand{\arraystretch}{1.6}
\Large
\resizebox{\textwidth}{!}{%
\begin{tabular}{|p{2.5cm}|p{2.5cm}|p{1.2cm}|p{2.3cm}|p{3cm}|p{3.8cm}|p{4.8cm}|p{6.1cm}|}
\hline
\textbf{Study} & \textbf{Species} & \textbf{Sample Size} & \textbf{Gestational Age} & \textbf{Signals Used} & \textbf{Method} & \textbf{States Identified} & \textbf{Performance} \\
\hline
Vairavan et al. (2016)~\cite{vairavan_computer-aided_2016} & Human fetuses & 39 & 30--38 weeks & FMCG (HR + Actogram) & Rule-based thresholds + ROC optimization & 
\begin{tabular}[t]{@{}l@{}}$<$36 wks: 1F vs. 2F \\ $\geq$36 wks: 1F vs. 2F\end{tabular} & 
\begin{tabular}[t]{@{}l@{}}$<$36 wks: ICC = 0.88 (1F), 0.65 (2F) \\ $\geq$36 wks: ICC = 0.88 (1F), 0.41 (2F) \\ AUC = 0.99 (both)\end{tabular} \\
\hline

Semeia et al. (2022)~\cite{semeia_evaluation_2022} & Human fetuses & 52 & 27--39 weeks & FMCG (HRV + Actogram) & Rule-based thresholds + ROC optimization & 
\begin{tabular}[t]{@{}l@{}}$<$32 wks: Active vs. Passive \\ $\geq$32 wks: 1F vs. 2F\end{tabular} & 
\begin{tabular}[t]{@{}l@{}}$<$32 wks: AUC $\approx$ 1.0 (HRV), \\0.80--0.83 (Actogram) \\ $\geq$32 wks: AUC $\approx$ 1.0 (HRV), \\0.86--0.87 (Actogram)\end{tabular} \\
\hline

Myers et al. (1993)~\cite{myers_quantitative_1993} & Fetal baboons & 3 & 143--153 days & EEG (frontal + parietal) & Rule-Based (K-means preprocessing) & 1F vs. 2F & 
\begin{tabular}[t]{@{}l@{}}Expert agreement: 87.1\% (Overall)\\ 79.7\% (1F), 91.3\% (2F)\end{tabular} \\
\hline

Grieve et al. (1994)~\cite{grieve1994behavioral} & Fetal baboons & 3 & 80\%--90\% of term & EEG, EOG, ECG & Rule-Based (K-means preprocessing) & 1F vs. 2F & 
\begin{tabular}[t]{@{}l@{}}Expert agreement: 81.5\% (Overall) \\ 83.7\% (1F), 79.4\% (2F)\end{tabular} \\
\hline

Samjeed (2022)~\cite{samjeed2022classification} & Human fetuses & 105 & 20--40 weeks & Non-invasive fetal ECG (NI-fECG) & 1D CNN & 1F vs. 2F & 
\begin{tabular}[t]{@{}l@{}}F1: 80.2\% (1F), 69.5\% (2F) \\ Accuracy: 76\% \\ Sensitivity: 72.7\% (1F), 82.6\% (2F)\end{tabular} \\
\hline

Subitoni (2022)~\cite{subitoni2022hidden} & Human fetuses & 115 & 27--39 weeks (grouped: early/mid/late) & FHR & HMM + CNN (Hybrid) & 1F vs. 2F & 
\begin{tabular}[t]{@{}l@{}}\textbf{HMM+CNN:} \\ F1: 87.87\%, Balanced Acc: 88.37\% \\ \textbf{HMM only:} \\ F1: 77.73\%, Balanced Acc: 83.30\%\end{tabular} \\
\hline
\end{tabular}
}
\end{table*}

\subsection{Fetal Heart Rate Variability Analysis}

\subsubsection{Physiological Basis of FHRV in Sleep}
Sleep states in fetuses are associated with distinct changes in physiological parameters, prominently fetal heart rate variability (FHRV) \cite{garland_fetal_nodate, shaw_altered_2018}. In humans, transitions between sleep states are reflected in changes in heart rate patterns, strongly correlating with behavioral states defined by heart rate, body movements, and eye movements \cite{nijhuis_are_1982}. Similar correlations have been observed in fetal baboons, where FHRV measures, combined with EOG and EEG data, have been successfully used to define behavioral state cycles \cite{grieve1994behavioral}. High EEG-Ratio periods, indicative of quiet sleep, correspond to lower heart rates and reduced FHRV in fetal baboons \cite{myers_quantitative_1993}, suggesting that ANS modulation of FHRV is influenced by sleep states \cite{grieve1994behavioral}.

In fetal sheep, physiological studies show distinct differences in FHRV between sleep states. Quiet sleep is typically associated with lower beat-to-beat variability compared to active sleep, indicating varying ANS modulation \cite{shaw_altered_2018}. 

\subsubsection{Feature Extraction for Sleep Classification}
Feature extraction from FHRV typically relies on time-domain and frequency-domain measures derived from RR intervals. In fetal baboons, features such as the standard deviation of RR intervals (SD-RR) and the root mean square of successive differences (RMSSD) are computed on a minute-by-minute basis, provided that at least 90\% of RR intervals are artifact-free \cite{garland_fetal_nodate}. In human fetal studies, SDNN, RMSSD, and permutation entropy have similarly been employed to classify sleep states \cite{brandle_heart_2015}. In fetal sheep, frequency-domain spectral measures—such as low-frequency (LF), HF, and the LF/HF ratio—have been used to differentiate FBS \cite{shaw_altered_2018}. However, the interpretation of LF/HF as a marker of sympatho-vagal balance is controversial, as LF power reflects a combination of sympathetic and parasympathetic influences, and the ratio can be affected by non-neural factors such as respiration and heart rate \cite{billman2013lf}. 

\subsection{Rule-based Approaches}
\subsubsection{Threshold-Based Classification Using FMCG and Actogram Signals}
Rule-based approaches for FBS classification typically rely on deterministic thresholds derived from physiological signals, such as HRV and actogram-based fetal movement data. These systems apply expert-defined rules to classify states by comparing extracted features with fixed thresholds. While these methods provide interpretable and practical solutions for assessing fetal sleep and wakefulness, they lack the flexibility to adapt to individual variability and gestational changes, limiting their robustness in real-world scenarios.



In 2016, Vairavan et al.~\cite{vairavan_computer-aided_2016} developed an early automated pipeline for FBS classification using FMCG recordings from 39 fetuses between 30 and 38 weeks of gestation. They translated Nijhuis criteria~\cite{nijhuis_are_1982} into fixed-threshold rules based on fetal heart rate patterns and actogram-derived movements. The system demonstrated strong agreement with expert annotations, particularly for quiet sleep (intraclass correlation coefficient (ICC) = 0.88), though performance declined for active sleep in later gestation (ICC dropped to 0.41). These findings suggest that while rule-based classification is feasible with FMCG and CTG, behavioral complexity increases with maturation, potentially limiting such approaches.



Building on Vairavan et al.'s work, Semeia et al.~\cite{semeia_evaluation_2022} refined rule-based FBS classification by introducing gestational age-specific distinctions. Using a large FMCG dataset, they separated younger ($<$32 weeks) and older ($\geq$32 weeks) fetuses and adapted the classification accordingly—distinguishing active/passive states in early gestation and 1F/2F states later. Their results showed that HRV-derived parameters, especially RMSSD and standard deviation (STD) of HR, achieved near-perfect classification accuracy (AUC $\approx$ 1.0), while actogram-based features were less reliable. These findings reinforce the utility of HRV metrics for FBS classification and highlight the need for developmental stage-specific models.

Both studies demonstrated that rule-based approaches can achieve high classification accuracy for prototypical FBS, but their reliance on fixed thresholds restricts their adaptability across different gestational ages and individual variations. The absence of a temporal component in these models makes it difficult to capture transitional states that naturally occur as fetal development progresses. Moreover, rule-based methods do not account for probabilistic uncertainty, potentially leading to overconfidence in misclassified instances.

Semeia et al.~\cite{semeia_evaluation_2022} identified a key limitation of rule-based methods: substantial overlap in parameters between quiet and active sleep, which hampers accurate classification—especially during transitional phases with gradual physiological changes. This suggests such methods may oversimplify the complex dynamics of FBS.

As shown in Table~\ref{tab:combined_fbs_comparison}, Vairavan et al.~\cite{vairavan_computer-aided_2016} and Semeia et al.~\cite{semeia_evaluation_2022} achieved good results using FMCG, but did not assess generalizability to more accessible modalities like CTG. Their methods also struggle with fetal state transitions and individual variability, highlighting the need for more advanced probabilistic and machine learning approaches.

\subsubsection{Rule-Based Classification Using K-means and EEG/Multimodal Signals}
Unsupervised clustering methods have been explored for FBS classification, particularly using K-means applied to spectral EEG features. Myers et al.~\cite{myers_quantitative_1993} proposed an early rule-based method using fetal baboon EEG data. They developed the “EEG-Ratio” defined as the power in the 0.03–0.2 Hz band (associated with trace alternant) divided by power in the 12–24 Hz band. This feature correlated with visually scored sleep states, and K-means clustering was applied to classify data into two binary states: trace alternant (TA, representing quiet sleep) and non-TA (active sleep). The resulting classification achieved an 87.1\% agreement with expert scoring.

However, the study had limitations. It included only three fetal baboons, each contributing four EEG recordings, totaling 3,694 minutes of usable data. The authors did not use any form of cross-validation or independent testing, as thresholds were optimized and validated on the same dataset, potentially inflating accuracy estimates. Moreover, the EEG-Ratio—being a scalar feature—may not generalize well across subjects or conditions.

To improve robustness, Grieve et al.~\cite{grieve1994behavioral} extended this approach by integrating multimodal signals—EEG, EOG, and ECG—from the same three fetal baboons, each recorded for 16 continuous hours. They applied K-means clustering to extract binary thresholds for three features: EEG ratio, EOG spectral power, and RR interval variability (CVRR). These features were then combined using rule-based criteria to define two sleep states: 1F (quiet sleep) and 2F (active sleep). Transitions and indeterminate states were also identified using temporal continuity rules. Agreement with expert annotations reached 81.5\% overall (83.7\% for 1F and 79.4\% for 2F), demonstrating the feasibility of long-term, automated multimodal classification.

While the use of multimodal features provided a more physiologically grounded framework, limitations remained, including small sample size, absence of gestational stratification, and no direct comparison with unimodal or machine learning-based models. Nonetheless, these early efforts laid the groundwork for automated fetal sleep state detection using interpretable, unsupervised approaches.

\subsection{Machine Learning Approaches}


Deep learning has emerged as a powerful tool for classifying FBS from physiological signals. Two recent studies—by Samjeed et al.~\cite{samjeed2022classification} and Subitoni et al.~\cite{subitoni2022hidden}—have proposed deep neural network-based methods leveraging fetal ECG and FHR signals, respectively.

Samjeed et al.~\cite{samjeed2022classification} proposed a 1D convolutional neural network (1D-CNN) to classify fetal behavioral states from non-invasive abdominal ECG recordings of 105 fetuses (20–40 weeks gestation, 3–10 min duration). The CNN consisted of three convolutional layers and was trained using stochastic gradient descent with momentum (SGDM) with 5-fold cross-validation. It achieved 76\% accuracy, with F1-scores of 80.2\% for the quiet state and 69.5\% for the active state, suggesting challenges in distinguishing between states. Limitations included dataset imbalance, lack of temporal modeling, and use of a single modality. Future directions included exploring recurrent neural networks (RNNs), multimodal inputs, and transfer learning.

Subitoni et al.~\cite{subitoni2022hidden} proposed a hybrid model combining hidden markov models (HMMs) and a U-Net style 1D-CNN (U-Sleep variant) to classify fetal behavioral states from 115 manually annotated FHR recordings. The dataset was stratified into three gestational age groups (27–32, 33–36, 37–39 weeks) to account for developmental differences.

The approach used HMMs for unsupervised segmentation to generate pseudo-labels, which were then used to pre-train a CNN. The model was subsequently fine-tuned using expert annotations. This two-stage training allowed the system to leverage both unlabeled and labeled data.

The hybrid model achieved a Macro F1-score of 87.87\%, outperforming the HMM alone (77.73\%). However, the study did not clarify whether evaluation was subject-wise or sample-wise, and relied on annotations from a single expert, limiting generalizability. Still, the method demonstrates a promising strategy to reduce dependence on annotated data via hybrid learning.

\section{Effects of Abnormal Conditions on Fetal Sleep}

\subsection{Hypoxia}

Hypoxia is a major disruptor of fetal sleep and neurodevelopment. Graded hypoxia experiments in fetal sheep have demonstrated significant disruptions in sleep states, including altered ECoG and behavioral activity \cite{koos_fetal_1987, shaw_altered_2018}. Koos et al.~\cite{koos_fetal_1987} reported that mild hypoxia did not alter the incidence of low-voltage ECoG activity, FBM, or REMs, whereas moderate and severe hypoxia markedly suppressed both FBM and REMs. A critical threshold was identified, with a reduction in arterial oxygen content of $2.00 \pm 0.23$ ml/dl associated with inhibition of both eye and breathing activity.

Recent studies have further characterized hypoxia-induced brain dysfunction through temporal assessments of EEG power and frequency recovery. In a fetal sheep model of asphyxia, prophylactic creatine supplementation significantly improved EEG recovery after umbilical cord occlusion (UCO), with higher power and faster restoration of physiologically organized frequencies, and reduced electrographic seizure burden. These effects were accompanied by reduced cortical cell death and white matter gliosis \cite{tran2025prophylactic}. Moreover, integration of low- and high-voltage EEG activity with nuchal EMG recordings provided detailed insights into sleep state reorganization under hypoxic stress.

In contrast, while magnesium sulfate attenuated gliosis and modestly improved myelin density in white matter tracts, it failed to improve EEG power, frequency, or sleep-state cycling in preterm fetal sheep exposed to hypoxia–ischemia. Neuronal and oligodendrocyte survival were similarly unaffected, suggesting limited functional neuroprotection despite some histological benefit \cite{galinsky2023magnesium}.


Prolonged hypoxia in late gestation can cause persistent suppression of EEG activity and dysmaturation of sleep states. In preterm fetal sheep, UCO led to a shift towards lower frequency EEG activity for the first five days, with a lasting reduction in EEG power in the delta and theta bands \cite{lear_dysmaturation_2024}. 
Chronic hypoxia impairs gestational age-related increases in overall fetal HRV, with evidence of suppressed sympathetic nervous system control of HRV after 72 hours of hypoxia exposure \cite{shaw_altered_2018}.  


Hypoxia alters fetal brain activity by promoting inhibitory neuromodulatory pathways, notably via elevated adenosine and increased neurosteroids such as allopregnanolone, resulting in predominant sleep-like EEG states \cite{efsa_panelon_animal_health_and_welfare_ahaw_animal_2017, yawno_neuroactive_2011}. In parallel with these central nervous system effects, hypoxia markedly reduces fetal forelimb movements and abolishes rapid eye movements, reflecting oxygen-conserving behavioral adaptations in fetal lambs \cite{natale_measurement_1981}. These EEG changes under hypoxic conditions are thought to result from adenosine-mediated inhibition and may suggest compensatory autonomic regulation~\cite{efsa_panelon_animal_health_and_welfare_ahaw_animal_2017}.
Acute fetal hypoxia, induced by reduced uterine blood flow or UCO, suppresses FBM, a response associated with elevated adenosine levels in the brain under hypoxic conditions \cite{koos1994hypoxic,koos1997source,watson2002effect}. Severe hypoxia further leads to muscle atonia, as reported by Breen et al. \cite{breen1997identification}. Neurophysiological studies indicate that the fetal midbrain regulates episodic breathing and mediates hypoxic inhibition of respiratory activity \cite{dawes1983breathing}. Koos et al.~\cite{koos1998thalamic} identified the parafascicular nuclear complex in the caudal thalamus as a critical structure mediating the suppression of fetal breathing during acute hypoxemia. While this inhibition is an acute response, it is important to recognize that severe or prolonged hypoxemia may lead to cerebral ischemia and encephalopathy, potentially altering fetal EEG patterns even after the hypoxic insult has passed~\cite{baburamani_brief_2021}.

 Hypercapnia during REM sleep enhances fetal breathing by increasing tracheal pressure and reducing apneic pauses, with CO$_2$ stimulation of breathing observed only during REM sleep in fetal lambs \cite{jansen_influence_1982}.
 Neurophysiological studies indicate that the fetal midbrain regulates episodic breathing and mediates hypoxic inhibition of respiratory activity \cite{dawes1983breathing}. Koos et al.~\cite{koos1998thalamic} identified the parafascicular nuclear complex in the caudal thalamus as a critical structure mediating the suppression of fetal breathing during acute hypoxemia. While this inhibition is an acute response, it is important to recognize that severe or prolonged hypoxemia may lead to cerebral ischemia and encephalopathy, potentially altering fetal EEG patterns even after the hypoxic insult has passed~\cite{baburamani_brief_2021}.

\subsection{Fetal Growth Restriction (FGR)}
FGR disrupts the normal organization of fetal sleep states, particularly in late gestation. Growth-restricted fetuses exhibit greater sleep-state instability, often spending more time in quiet sleep than in active sleep~\cite{arduini1989behavioural,van1985behavioural}. In addition, FGR is commonly associated with impaired oxygenation, reduced breathing and general movements, and an increased number of heart rate decelerations, reflecting ANS dysfunction \cite{cerritelli_role_nodate,bekedam1991effects}.

FGR fetuses also show diminished motor activity, characterized by slower, monotonous, and lower-amplitude movements, which reflect central nervous system impairment \cite{reissland_fetal_2016,bekedam1985motor,sival1992effect}. These disturbances are considered late-stage indicators of fetal compromise, often preceded by abnormalities in HRV and blood flow parameters \cite{reissland_fetal_2016,visser_fetal_2010}. Collectively, these findings highlight FBS as important indicators of neurodevelopment and fetal well-being.

\subsection{Fetal Congenital Malformations}
Structural or chromosomal abnormalities often correlate with altered fetal sleep states and reduced fetal movements (hypokinesia), linked to prolonged periods of low heart rate variability \cite{de2007changes,reissland_fetal_2016,visser_fetal_2010}.
\subsection{Other Maternal Conditions}
Several maternal and fetal abnormalities influence fetal sleep patterns, directly or indirectly:

\subsubsection{Maternal Diabetes}
Fetuses of diabetic mothers may show delayed sleep-state development between 32 and 40 weeks, often lacking the typical increase in quiet and active sleep seen in normal pregnancies~\cite{dierker1982change}.

\subsubsection{Maternal Sleep Position}
In the third trimester, the maternal supine sleep position significantly impacts FBS, increasing the likelihood of a transition towards quiet sleep due to reduced uterine perfusion and oxygen availability \cite{stone_effect_2017}. Such adaptation might reflect a fetal compensatory mechanism for reduced oxygen supply \cite{stone_effect_2017,stone_investigation_2017}.

 \subsubsection{Maternal Anxiety and Stress}

 Although explicit data on fetal sleep states remain limited, growing evidence highlights the significant impact of maternal psychological stress and anxiety on fetal development. Maternal self-reported anxiety has been associated with reduced expression of placental 11$\beta$-hydroxysteroid dehydrogenase type 2 (11$\beta$-HSD2), an enzyme critical for inactivating cortisol, thereby increasing fetal exposure to maternal glucocorticoids and potentially disrupting fetal neuroendocrine development~\cite{o2012maternal}. Such endocrine alterations represent one of the physiological pathways suspected to contribute to the later emergence of psychological disorders~\cite{opler2005fetal}.

 Longitudinal studies have elucidated the role of the pregnancy period in the intergenerational transmission of stress~\cite{lehrner2014maternal}. Prenatal stress exposure has also been associated with impaired fetal growth~\cite{spann2020association}, and these growth alterations are further linked to subsequent behavioral traits, including temperament in childhood~\cite{dipietro2018predicting} and adolescence~\cite{schlotz2014prenatal}.

 In addition to growth-related outcomes, maternal stress has been shown to directly alter the development of the fetal ANS. Specifically, maternal stress can entrain FHR patterns with maternal heart rate (HR) decelerations during respiratory efforts~\cite{lobmaier2020fetal}, and stressed mothers exhibit altered maternal-fetal HR coupling, characterized by significant decreases in FHR, suggesting the presence of a fetal stress memory that may serve as a novel non-invasive biomarker of prenatal stress exposure~\cite{lobmaier2020fetal}. Furthermore, recent evidence suggests that placental calcifications may reflect cumulative prenatal exposure to maternal stress and disease, potentially acting as a biological memory of the intrauterine environment. These placental adaptations may influence offspring cardiovascular and metabolic health through modulation of fetal ANS development~\cite{wallingford2018placental}.

 Collectively, these findings imply that maternal stress during pregnancy may have broad implications for fetal neurobehavioral development, including pathways potentially relevant to the regulation of sleep states, autonomic function, and long-term health outcomes.

\subsubsection{Alcohol Consumption}
Alcohol exposure, even episodic, disrupts fetal REM sleep and drastically reduces fetal breathing activity, potentially leading to long-term neurobehavioral consequences characteristic of fetal alcohol syndrome (FAS)~\cite{mulder1998acute}. Such effects were demonstrated in a controlled study where maternal consumption of two glasses of wine suppressed fetal REM activity and breathing movements~\cite{mulder1998acute}. These disruptions may underlie some of the neurobehavioral and ophthalmic deficits observed in FAS~\cite{reissland_fetal_2016}.

 \subsubsection{Smoking}
 Smoking during pregnancy has been associated with significant alterations in PE, reflecting disrupted autonomic regulation before birth \cite{zeskind2006maternal, kapaya2015smoking, mulkey2019autonomic}. Such autonomic disruptions may also suggest potential disturbances in fetal sleep-state organization, though specific sleep-related effects require further investigation.

\subsubsection{Caffeine Intake}
Maternal caffeine consumption affects FBS organization, increasing general body movements and altering sleep-wake patterns, with a trend toward reduced breathing activity \cite{mulder_foetal_2010}.

\subsubsection{Magnesium Sulfate Administration}
Magnesium sulfate reduces FBM \cite{han_fetal_2018} and has been reported to cause fetal bradycardia and diminished HRV, possibly through maternal hypothermia or direct fetal cardiac effects~\cite{cardosi1998magnesium}. Recent studies in preterm fetal sheep further demonstrate that magnesium sulfate suppresses FHR, EEG activity, and increases cardiac afterload, all of which may influence fetal behavioural states \cite{galinsky2016magnesium,galinsky2017magnesium,galinsky2018magnesium}. In addition, sex-specific differences have been observed in EEG suppression and cardiovascular responses to hypoxia, suggesting that fetal sex may modulate the effects of pharmacological interventions on fetal behaviour \cite{galinsky2018magnesium}. Additionally, concurrent use of magnesium sulfate and nifedipine may result in severe hypotension and cardiac depression in the fetus \cite{khedun2000effects}. These effects may potentially impact FBS, although specific disruptions in sleep-related behaviors remain to be clarified.

\subsubsection{Antidepressants}
Selective serotonin reuptake inhibitors are commonly prescribed to manage maternal anxiety and depression during pregnancy~\cite{hanley2016fetal}.

According to findings reported by Mulder et al.~\cite{mulder_selective_2011} and summarized in this chapter~\cite{reissland_fetal_2016}, fetal exposure to standard or high Selective serotonin reuptake inhibitors dosages was associated with increased general movements and disrupted NREM sleep near term, characterized by persistent bodily activity and impaired inhibitory motor control during quiet states. However, the significance of poor fetal sleep regulation for postnatal neurobehavioral development remains unclear and warrants further investigation~\cite{reissland_fetal_2016}.

\subsection{Intra-amniotic Infection (Chorioamnionitis)}
Intra-amniotic infections, including clinical and subclinical chorioamnionitis, have been associated with loss of fetal heart rate cycling and adverse perinatal outcomes, supporting the role of inflammation in fetal behavioral dysregulation \cite{galli2019intrapartum}. More recently, absence of fetal heart rate cycling has also been linked to maternal intrapartum pyrexia, with affected fetuses showing lower neonatal Apgar scores, further suggesting disruption in FBS organization \cite{pereira_absence_2025}. As fetal heart rate cycling reflects the alternation between active and quiet sleep, these findings imply that intra-amniotic infection may disturb fetal sleep-wake cycling. Consistent with this, progressive systemic inflammation in late-gestation fetal sheep leads to suppression of high-frequency EEG activity, particularly in the beta and gamma bands, which are believed to reflect cortical activation and sleep state transitions. These EEG alterations were sustained even after the resolution of inflammation, suggesting persistent disruption of fetal behavioral state cycling \cite{kelly2023progressive}.


In conclusion, although specific evidence on the direct impact of these abnormalities on fetal sleep is limited, the observed disruptions in fetal movements, HRV and breathing patterns highlight potential implications for fetal sleep states. Understanding these links further emphasizes the importance of monitoring fetal sleep as a critical parameter in fetal surveillance.


\section{Future Directions and Challenges}

\subsection{Limitations of Current Studies}

\subsubsection{Technological Limitations}
While current tools have advanced FBS research, technologies like ultrasound, CTG, and FMCG face limitations such as low signal quality, motion artifacts, and limited applicability in early gestation. Fetal MRI provides better CNS insights but is costly and inaccessible. Overall, existing methods lack the resolution and scope to fully capture early fetal neurodevelopment and sleep transitions.

\subsubsection{Analytical Challenges}
FBS detection often depends on visual inspection or algorithms trained on prototypical segments, overlooking transitional or ambiguous states that may offer important developmental insights. Inconsistent terminology across studies further hinders reproducibility and comparison. Most methods also fail to capture complex dynamics like diurnal rhythms or maternal-fetal interactions.

\subsubsection{Sample Size and Interindividual Variability}
Small sample sizes limit the statistical power and generalizability of fetal sleep studies, especially in linking FHR to biochemical markers. Variability across fetuses—driven by gestational age, maternal health, and environmental factors—complicates standardizing FBS classification. Broad gestational groupings (e.g., mid and late gestation) may mask critical developmental transitions.

\subsubsection{Longitudinal and Genetic Considerations}
Links between prenatal sleep and postnatal outcomes are limited by long assessment gaps and unaccounted genetic influences shared by mother and fetus. This highlights the need for integrated, genetically-informed longitudinal studies to better explain outcome variability.

\subsection{Future Research Directions}
To overcome current limitations, future research should explore advanced analytical tools—such as point process models and detailed HRV metrics—to uncover biomarkers of fetal brain and ANS development. Although fetal EEG is infeasible in humans, invasive recordings in animal models (e.g., sheep, baboons) can inform the interpretation of non-invasive human data (e.g., FMEG, FMCG, coherence). Cross-modal integration of spatio-temporal and synchrony features may further elucidate fetal CNS maturation and sleep-state transitions.

Improving automated FBS detection remains critical. Machine learning models, validated against tools like fetal MRI and synchronized physiological signals, can increase reproducibility and standardization. Establishing consistent terminology and definitions will also enhance model generalizability and cross-study comparability.

To address small sample sizes and individual variability, transfer learning is a promising solution. Pretrained models on adult sleep data can be fine-tuned on fetal recordings, leveraging shared low-level features while adapting high-level patterns to fetal physiology. This reduces data demands and improves model robustness across gestational ages and maternal-fetal conditions.
In addition, adult sleep data can be transformed to better match the spectral characteristics of fetal sleep data using signal processing or generative adversarial networks (GANs). Aligning spectral distributions in this way provides a more compatible source for fine-tuning, further enhancing transfer learning performance. Inspired by reinforcement learning, reward-guided fine-tuning based on physiological plausibility or expert preference can further improve adaptation across gestational stages.

A multidisciplinary approach—linking neuroscience, obstetrics, and neuroimaging—is essential. Longitudinal studies from early gestation to infancy with continuous maternal-fetal monitoring can clarify developmental trajectories, especially sleep-state transitions and maternal influences.

Further exploration of vagal tone and ANS maturation may identify critical periods of vulnerability. Refining measurement tools through multimodal integration will be key to developing clinical guidelines for identifying fetuses at risk of autonomic or neurodevelopmental disorders.

Ultimately, a comprehensive perinatal perspective—recognizing bidirectional maternal-fetal interactions and the continuity of sleep-state development—is vital. Monitoring fetal sleep may enable early detection of FGR, chronic hypoxia, or emerging neurological issues. Early intervention (e.g., optimized delivery, neuroprotective agents, maternal care) is crucial for improving long-term outcomes during this sensitive developmental window.

\bibliographystyle{ieeetr}  %
\bibliography{reference}    

\end{document}